\documentclass[10pt, twocolumn, nofootinbib,superscriptaddress]{revtex4-1}
\usepackage{amsmath,amssymb,amsfonts}
\usepackage{algorithmic}
\usepackage{graphicx}
\usepackage{textcomp}
\usepackage{xcolor}
\usepackage{ragged2e}
\usepackage{booktabs, makecell, tabularx}
\usepackage{lipsum}

\usepackage{gensymb}
\usepackage{multirow}
\makeatletter
    \renewcommand\@make@capt@title[2]{%
     \@ifx@empty\float@link{\@firstofone}{\expandafter\href\expandafter{\float@link}}%
      {\textbf{#1}}\@caption@fignum@sep#2\quad}%
\makeatother
\makeatletter 
\renewcommand{\fnum@figure}{\textbf{Fig.~\thefigure}} 
\makeatother

\usepackage{xcolor}
\newcommand{\beginsupplement}{%
        \setcounter{table}{0}
        \renewcommand{\thetable}{S\arabic{table}}%
        \setcounter{figure}{0}
        \renewcommand{\thefigure}{S\arabic{figure}}%
     }
\def\BibTeX{{\rm B\kern-.05em{\sc i\kern-.025em b}\kern-.08em
    T\kern-.1667em\lower.7ex\hbox{E}\kern-.125emX}}

\begin{document}

\author{Zheng Zheng}
\affiliation{Nonlinear Nanophotonics Group, MESA+ Institute of Nanotechnology,\\
University of Twente, Enschede, The Netherlands}
\author{Ahmet Tarık Işık}
\affiliation{Nonlinear Nanophotonics Group, MESA+ Institute of Nanotechnology,\\
University of Twente, Enschede, The Netherlands}
\author{Akshay Keloth}
\affiliation{Nonlinear Nanophotonics Group, MESA+ Institute of Nanotechnology,\\
University of Twente, Enschede, The Netherlands}
\author{Kaixuan Ye}
\affiliation{Nonlinear Nanophotonics Group, MESA+ Institute of Nanotechnology,\\
University of Twente, Enschede, The Netherlands}
\author{Peter van der Slot}
\affiliation{Nonlinear Nanophotonics Group, MESA+ Institute of Nanotechnology,\\
University of Twente, Enschede, The Netherlands}
\author{David Marpaung}
\email{david.marpaung@utwente.nl}
\affiliation{Nonlinear Nanophotonics Group, MESA+ Institute of Nanotechnology,\\
University of Twente, Enschede, The Netherlands}

\date{\today}
\title{Thermoelastic surface acoustic waves in low-loss silicon nitride integrated circuits}

\begin{abstract}
Acousto-optic modulation in photonic integrated circuits harness the applications that include signal processing, quantum photonics and microwave photonics. However, silicon nitride ($\rm{Si_3N_4}$), as a main-stream low-loss scalable photonic platform, suffers from the lack of piezoelectric effect and therefore the hybrid co-integration with other materials is always required for acousto-optic modulation. Here, we employed thermoelastic surface acoustic waves (SAW) in a 8 dB/m propagation loss $\rm{Si_3N_4}$ integrated circuits without adding extra materials. A phase modulation efficiency enhancement of 13.6 dB is realized with a multi-pass configuration. Furthermore, a single-sideband intermodal scattering with a suppression ratio of 8~dB is measured and an intensity modulation is observed by incorporating the phase modulation into a ring resonator spectral. This thermoelastic SAW technique, as an initial step of acousto-optic modulation in low-loss $\rm{Si_3N_4}$ platform, is promising for integrated microwave photonics and programmable photonics applications. 
\end{abstract}

\maketitle
\section{Introduction}
Introducing acoustic waves into optical platforms provides a mechanism for active functions in integrated photonic devices. The acoustic waves could modulate the optical refractive index of the materials by inducing a distributed grating structure. This acousto-optic effect has been incorporated in many important platforms, for example, silicon (Si) \cite{Zhou2024,Ansari2022,Huang2022,Kittlaus2021}, lithium niobate (LN) \cite{Cai2019,Yu2021,Wan2022,Yu2024,Shao2020, Ye2025}, and aluminum nitride (AlN) \cite{Liu2019}. By integration with other photonic functionalities, the acousto-optic effect can be tailored for applications in microwave photonics \cite{Marpaung2019,Wei2025,Feng2024,Ye2025}, quantum photonics \cite{Ansari2022,Raphael2015} and data communication \cite{Zhou2024,Kittlaus2021,Cai2019,Yu2024,Shao2020, Wan2022, Balram2016}. 

Silicon nitride ($\rm{Si_3N_4}$) is a mature photonic integration platform  with the benefits of ultra-low propagation loss \cite{Bose2024}, wide transparency window \cite{Xiang2022}, high optical power handling capability \cite{Ming2010}, and has a fabrication process that is compatible with the commercial CMOS foundry process \cite{Freedman2025}. Therefore, $\rm{Si_3N_4}$ has been widely explored in Kerr frequency comb \cite{Kim2017}, sensing \cite{Zhou2022sci}, quantum physics \cite{AghaeeRad2025} and so forth. However, $\rm{Si_3N_4}$ is a naturally passive material \cite{Xiang2022} without piezoelectric effect and a low photoelastic coefficient compared to Si or LN \cite{Botter2022}, so the modulator demonstration is restricted up to tens of kHz bandwidth \cite{Nejadriahi2021,Alemany2021} by thermo-optic effect. Although by heterogeneous integration through additional deposition \cite{Kittlaus2021,Ansari2022,Huang2022,Zhou2024,Kenning2025,Freedman2025} or by wafer bonding  \cite{Snigirev2023}, active materials, such as III-V semiconductor or piezo-active materials, could be placed on top of $\rm{Si_3N_4}$ to introduce active components, but this typically compromises the optical propagation loss. In order to break this fundamental limitation of $\rm{Si_3N_4}$, we introduce the thermoelastic mechanism to generate acoustic waves on a low-loss $\rm{Si_3N_4}$ photonic device. 

\begin{figure*}[htp]
\centering
\includegraphics[width=\linewidth]{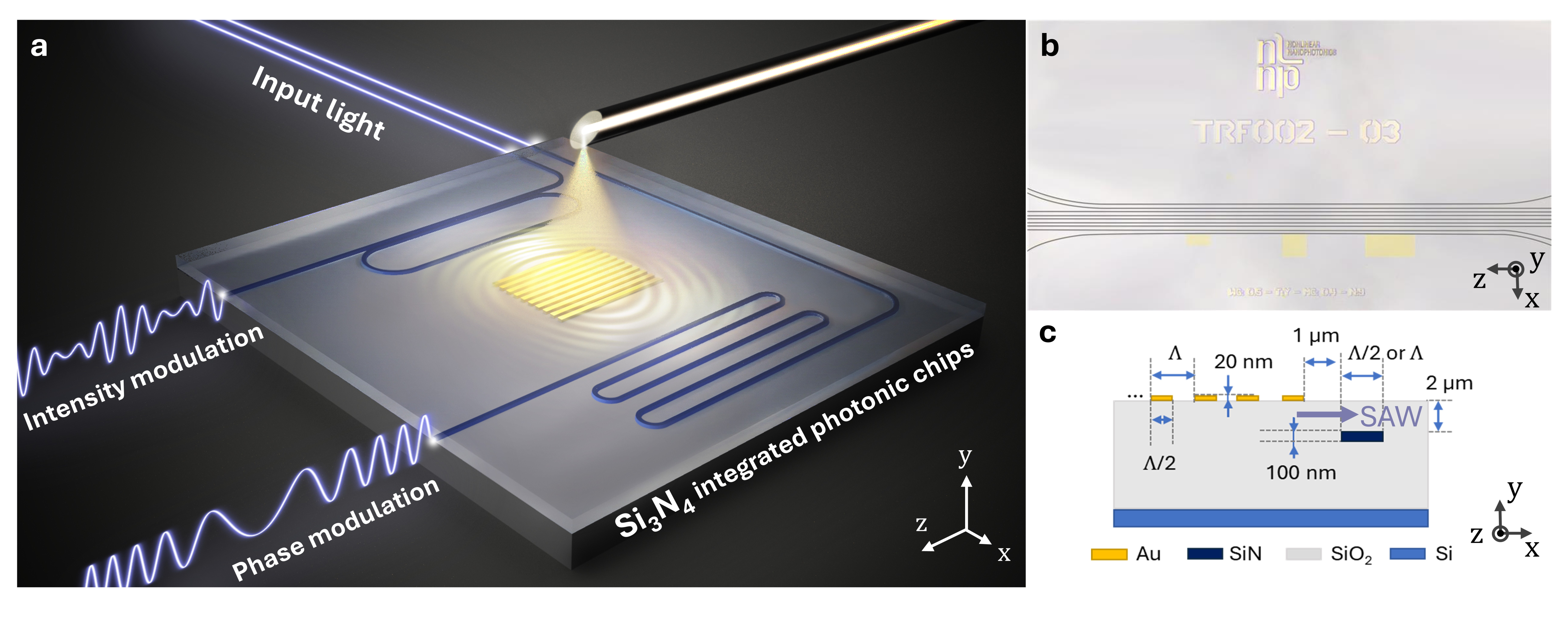}
\caption{(a) Conceptual figure. A gold metallic grating is illuminated by an intensity-modulated pump light. The thermoelastic SAW interacts with optical waves in the low-loss $\rm{Si_3N_4}$ waveguides. (b) Microscope photo of the phase modulation design. The black lines are multiple waveguide-passes and the yellow blocks are metallic gratings. (c) Chip cross-section view. The dimension is not to scale.}
\label{fig-concept}
\end{figure*}

Thermoelastic surface acoustic waves (SAW) are excited via absorption of an intensity-modulated pump light in a metallic grating, triggering periodic elastic displacement in the underlying material (see Figure~\ref{fig-concept}(a)). This was first reported in bulk crystalline quartz and Si \cite{Bonello2001}. After almost two decades, it was revived in integrated photonics form and applied to the silicon on insulator (SOI) platform \cite{Munk2019,Kittlaus2021}. Recently, this technique was successfully adapted in $\rm{Si_3N_4}$ \cite{Shafir2025,Zheng2025-cleo} and thin-film lithium niobate (TFLN) \cite{Zheng2025} platforms. Particularly in these recent reports, thermoelastic SAW was generated in the silica cladding, which is the most common cladding material of integrated photonic devices. While currently the modulation efficiency of thermoelastic SAW is still lower than that of electrically-driven devices, it benefits from flexible choice in integrated platform, high integration, programmable photonic functions, and free of impedance-matching design. 

In this paper, we report on thermoelastic SAW in a low-loss $\rm{Si_3N_4}$ integrated photonic device with three different demonstrations while maintaining a low optical propagation loss of 8~dB/m. First, we demonstrate efficient acousto-optic phase modulation where the low propagation loss enables long interaction length of light and the acoustic waves through multiple waveguide-passes, which results in a 13.6-dB modulation efficiency enhancement at a modulation frequency of 0.81~GHz using a gold metallic grating with a 4-$\mu$m period. Secondly, We demonstrate a single-sideband acousto-optic modulation through intermodal scattering at 0.76~GHz with a 6-$\mu$m period metallic grating, achieving an 8-dB extinction. Finally, we demonstrate a conversion of phase-to-intensity modulation using a ring resonator, observing intensity modulation signals at two frequencies of 0.95 and 1.75~GHz. Leveraging these functions, this technique adds more functions to integrated microwave photonics and programmable photonics in $\rm{Si_3N_4}$ platform.
\bigskip

\section{Results}
\textbf{Concept and chip design} The conceptual figure is shown in Figure~\ref{fig-concept}(a). The pump light is delivered from a polished fiber with a 40$\degree$ end-facet and then shines on a gold metallic grating. The $\rm{Si_3N_4}$ core waveguides are positioned closely to the metallic grating to enable the overlap between optical waves and acoustic waves. To enhance the modulation efficiency, a multi-pass configuration is utilized to let the SAW pass through the waveguides several times. A microscope photo of the chip top-view is captured in Figure~\ref{fig-concept}(b) and the chip cross-section view in Figure~\ref{fig-concept}(c) indicates the dimensions of the layer structure. The 100-nm thick low-loss $\rm{Si_3N_4}$ core waveguide is covered by a 2-$\mu m$ thick silica top cladding. The 20-nm thick gold metallic grating acts as an acoustic emitter on the surface. For phase modulation and intensity modulation, the waveguide width is half of the grating period, and for intermodal scattering, they are equal.

\begin{figure*}
\centering
\includegraphics[width=\linewidth]{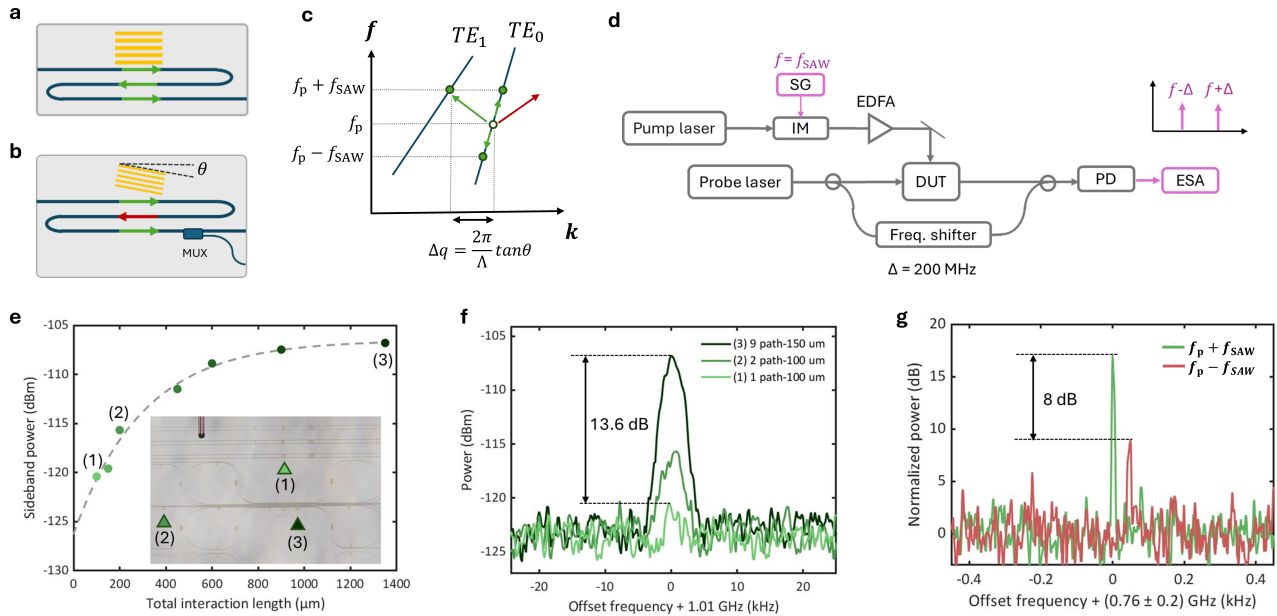}
\caption{Phase modulation and intermodal scattering configurations. (a) A 4-$\mu$m period metallic grating is placed parallel to a multiple waveguide-passes. Each path constructively contributes to the modulation efficiency. (b) A 6-$\mu$m period metallic grating is titled at an angle $\theta$ with respect to the waveguides to match the wavevetor difference between $\rm{TE_0}$ and $\rm{TE_1}$ optical modes. However, only the forward path (green arrow) matches, the backward path (red arrow) does not. (c) Phase-matching plots of phase modulation and intermodal scattering processes. (d) The heterodyne measurement setup. The intensity-modulated pump light is amplified and then delivered to the metallic grating with a 40\degree facet-polished fiber. The probe light is partially guided to the device, and the rest to an frequency shifter. After passing the device, they combine and beat on the PD. The beating signals are observed on ESA. (e) The measured sidebands power on ESA vs. the total interaction length. (f) A modulation efficiency enhancement of 13.6 dB with a 9-path configuration. (g) The intermodal scattering performs a 8-dB suppression ratio between two sidebands. IM: intensity modulator, SG: signal generator, EDFA: erbium-doped fiber amplifier, DUT: device under test, PD: photo detector, and ESA: electrical spectrum analyzer. }
\label{fig-pm-im}
\end{figure*}

\textbf{Phase modulation}. By positioning the metallic grating close and parallel to the waveguides, as illustrated in Figure~\ref{fig-pm-im}(a), we demonstrate the phase modulation. The phase-matching plot in Figure~\ref{fig-pm-im}(c) shows, under the phase-matching condition, the phase modulation generates two sidebands at $f_{\rm{p}} \pm f_{\rm SAW}$. We build a heterodyne measurement setup, on Figure~\ref{fig-pm-im}(d), to characterize the signals \cite{Zheng2025}. We start with an external intensity modulator (IM) to modulate the pump light intensity, and following an erbium-doped fiber amplifier (EDFA) to amplify the average off-chip pump optical power to 670~mW. For the detection of phase modulation signals, the heterodyne measurement setup downshifts the modulated optical frequencies to radio frequency, and is finally observed on an electrical spectrum analyzer (ESA). With the multi-pass configuration, a modulation efficiency enhancement of 13.6~dB is observed with up to 9 paths, as shown in Figure~\ref{fig-pm-im}(e-f). This enhancement approaches a saturation point is because of the SAW dissipation during propagation. 

\textbf{Intermodal scattering}. To achieve intermodal scattering, we fabricate the gratings with a slight tilt angle with respect to a 3-$\mu$m width multimode waveguide, as shown in Figure~\ref{fig-pm-im}(b). The metallic grating is placed at an angle $\theta$ with respect to the optical waveguide, so that the acoustic wave vector compensates the wave vector differences between $\rm{TE_0}$ and $\rm{TE_1}$ optical modes, $\Delta q = 2 \pi tan \theta / \Lambda$. Therefore, in the forward propagation direction, the acoustic waves scatter the light from fundamental $\rm{TE_0}$ mode to higher-order $\rm{TE_1}$ mode, and shifts the frequency from $f_{\rm{p}}$ to $f_{\rm{p}}+f_{\rm{SAW}}$. However, in the backward propagation direction, the phase matching condition is no longer satisfied, as indicated in Figure~\ref{fig-pm-im}(c). The optical waves in the antisymmetric mode is coupled out with a mode (de)multiplexer (MUX). We use the same measurement setup as Figure~\ref{fig-pm-im}(d), and the measured single-sideband suppression ratio is 8 dB shown in Figure~\ref{fig-pm-im}(g) at 0.76~GHz with a 6-$\mu$m period metallic grating. The remaining lower sideband is generated from phase modulation, because the phase mismatch here, $\Delta q L \ll 1$, can be neglected for phase modulation, where $L =$ 100 $\mu m$ is the acousto-optic interaction length.

\begin{figure*}
\centering
\includegraphics[width=\linewidth]{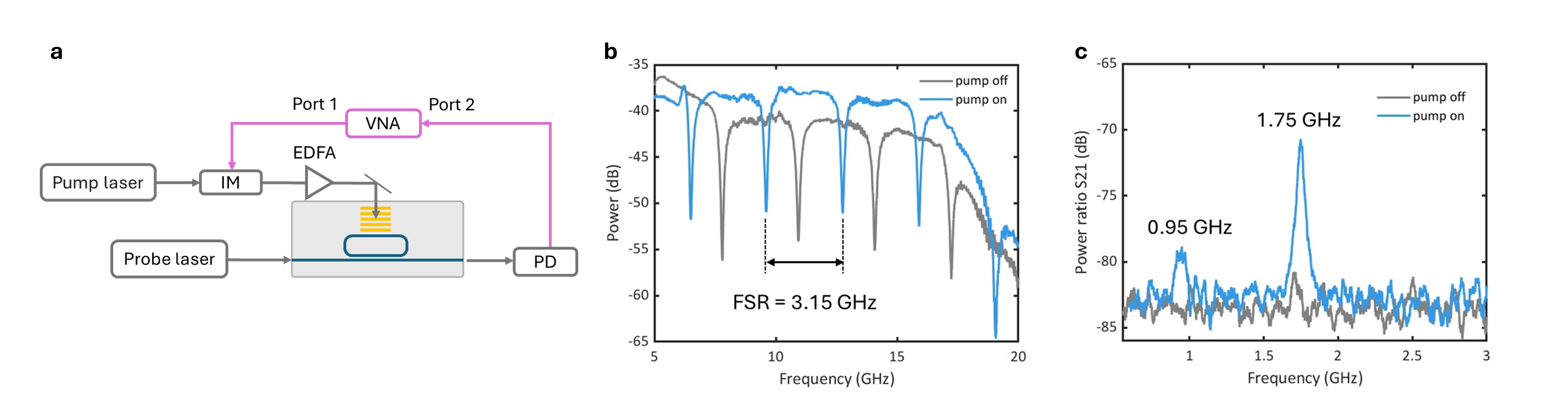}
\caption{Intensity modulation on a ring resonator. (a) The experimental setup. (b) Transmission spectrum of the ring resonator when turning on and off the EDFA. (c) The vector network analyzer (VNA) S21 measurement result. Two signals at 0.95 GHz and 1.75 GHz are generated from one of the fundamental modes of SAW and a second-harmonic generation of another fundamental mode, respectively. FSR: free spectral range.}
\label{fig-ring}
\end{figure*}

\textbf{Phase-to-intensity modulation conversion}. We also demonstrate intensity modulation by positioning the modulated sidebands at the spectral slop of a ring resonator transmission function, to transfer the phase modulation to intensity modulation. Figure~\ref{fig-ring}(a) shows the intensity modulation characterization experiment setup using a vector network analyzer (VNA). The ring resonator transmission spectrum is indicated on Figure~\ref{fig-ring}(b), with a free spectral range (FSR) of 3.15~GHz. We observe a spectrum shift because of the heating effect when the pump light shines on the metallic grating. The measurement result on Figure~\ref{fig-ring}(c) indicates the radio frequency power difference between two ports of VNA (S21). By tuning the probe light at the spectral slope of one of the ring responses, the measured signal at 0.95~GHz is one of the fundamental SAW modes, and signal at 1.75~GHz is the second-harmonic generation of another fundamental mode. By shaping the ring FSR and probe wavelength position, the ratio between fundamental and second-harmonic modulation frequencies can be engineered to target applications. 
 
\section{Discussion}
In this work, we use an intensity-modulated pump light to excite thermoelastic SAW, and report three different demonstrations to measure the acousto-optic interactions. Since the $\rm{Si_3N_4}$ waveguides have a low propagation loss of 8 dB/m, we design the multi-pass configuration to extend the acousto-optic interaction length and realize an enhancement of 13.6~dB on phase modulation efficiency. The refractive index perturbation is estimated to $\rm{3\times10^{-8}}$ refractive index units. Furthermore, we synchronize it with phase-matched optical modes and ring resonators to demonstrate single-sideband intermodal scattering and intensity modulation, respectively. More importantly, this thermoelastic acousto-optic modulation is realized solely on $\rm{Si_3N_4}$, without adding extra layers of materials, and preserving the low-loss property after adding the metallic gratings as acoustic emitters. 

In the future works, it is necessary to engineer the cladding thickness, metallic gratings pattern and integrate the pump delivery process within the chip with grating couplers, to further enhance the modulation efficiency. Nevertheless, this initial validation of thermoelastic SAW and its three direct acousto-optic modulation demonstrations represent the potential opportunities for $\rm{Si_3N_4}$ integrated circuits in programmable photonic applications. 

\section{Methods}

\textbf{Fabrication of devices}. A 100-nm thick $\rm{Si_3N_4}$ layer was deposited via low-pressure chemical vapor deposition (LPCVD) onto a silicon wafer with 8 $\mu$m thermal oxide, followed by annealing at 1200 $^\circ$C. Waveguides were patterned using electron beam lithography (EBL) and transferred to the silicon nitride layer via reactive ion etching (RIE). Subsequently, a 2-$\mu$m thick $\rm{SiO_2}$ layer was deposited via LPCVD. Metallic gratings were defined by EBL, followed by magnetron sputtering of a 5 nm Cr adhesion layer and a 20 nm Au layer. Finally, lift-off was performed in acetone under ultrasonication.

\textbf{Experimental details}. The pump laser is a single-frequency laser at 1558 nm (optilab, ULDC-1550-
MC) and the probe laser is a tunable laser at around 1557 nm (Thorlabs, DFB15TK). The pump light is modulated with an intensity modulator (Thorlabs, LN81S-FC) driven by a signal generator (Wiltron, 69147A). After modulation, pump light is amplified by an EDFA (Amonics, AEDFA-C). The heterodyne measurement is between the device output and the output from a frequency shifter (Aerodiode, RFAOM-AT-200) and is collected by a photodiode (PD, optilab, PD-23-C-DC) and shown on ESA (Keysight, N9000B). The intensity modulation measurement utilizes the VNA (Keysight, P5005A-200, 2-port). 

\section*{Author Contribution}
Z.Z. and D.M. proposed the concept. Z.Z., T.I. and A.K. designed and fabricated the silicon nitride photonic circuits. Z.Z. developed and performed the simulations and the experiments with input from D.M. and P.S.. Z.Z., P.S. and D.M. wrote the manuscript with input from all authors. D.M. led and supervised the entire project. 
\label{sec:four}
\bigskip

\section*{Acknowledgments}
The authors acknowledge funding from the European Research Council Consolidator Grant (101043229
TRIFFIC), Nederlandse Organisatie voor Wetenschappelijk Onderzoek (NWO) Start Up (740.018.021), Photon Delta National Growth Fund programme, the Research Grants Council.

\section*{Disclosures}
The authors declare no conflicts of interest.

\section*{Data Availability}
The data of this study are available
from the corresponding authors upon reasonable request.

\bibliographystyle{IEEEtran}
\bibliography{SAW-Arxiv-Ref}

@article{Munk2019,
   author = {Dvir Munk and Moshe Katzman and Mirit Hen and Maayan Priel and Moshe Feldberg and Tali Sharabani and Shahar Levy and Arik Bergman and Avi Zadok},
   doi = {10.1038/s41467-019-12157-x},
   journal = {Nat. Commun.},
   month = {4214, },
   publisher = {Nature Publishing Group},
   title = {Surface acoustic wave photonic devices in silicon on insulator},
   volume = {10},
   year = {2019},
}

@article{Liu2019,
   author = {Qiyu Liu and Huan Li and Mo Li},
   doi = {10.1364/optica.6.000778},
   isbn = {9781943580644},
   number = {6},
   journal = {Optica},
   publisher = {Optica Publishing Group (formerly OSA)},
   title = {Electromechanical Brillouin scattering in integrated optomechanical waveguides},
   volume = {6},
   pages = {778-785},
   year = {2019},
}

@article{Cai2019,
   author = {Lutong Cai and Ashraf Mahmoud and Msi Khan and Mohamed Mahmoud and Tamal Mukherjee and James Bain and Gianluca Piazza},
   doi = {10.1364/prj.7.001003},
   issn = {23279125},
   number = {9},
   journal = {Photonics Res.},
   pages = {1003-1013},
   publisher = {The Optical Society},
   title = {Acousto-optical modulation of thin film lithium niobate waveguide devices},
   volume = {7},
   year = {2019},
}

@article{Kittlaus2021,
   author = {Eric A. Kittlaus and William M. Jones and Peter T. Rakich and Nils T. Otterstrom and Richard E. Muller and Mina Rais-Zadeh},
   doi = {10.1038/s41566-020-00711-9},
   issn = {17494893},
   journal = {Nat. Photonics},
   pages = {43-52},
   publisher = {Nature Research},
   title = {Electrically driven acousto-optics and broadband non-reciprocity in silicon photonics},
   volume = {15},
   year = {2021},
}

@article{Wan2022,
   author = {Lei Wan and Zhiqiang Yang and Wenfeng Zhou and Meixun Wen and Tianhua Feng and Siqing Zeng and Dong Liu and Huan Li and Jingshun Pan and Ning Zhu and Weiping Liu and Zhaohui Li},
   doi = {10.1038/s41377-022-00840-6},
   issn = {20477538},
   issue = {1},
   journal = {Light Sci. Appl.},
   month = {145, },
   publisher = {Springer Nature},
   title = {Highly efficient acousto-optic modulation using nonsuspended thin-film lithium niobate-chalcogenide hybrid waveguides},
   volume = {11},
   year = {2022},
}

@article{Yu2021,
   author = {Zejie Yu and Xiankai Sun},
   doi = {10.1021/acsphotonics.0c01607},
   issn = {23304022},
   number = {3},
   journal = {ACS Photonics},
   pages = {798-803},
   publisher = {American Chemical Society},
   title = {Gigahertz Acousto-Optic Modulation and Frequency Shifting on Etchless Lithium Niobate Integrated Platform},
   volume = {8},
   year = {2021},
}

@article{Shao2020,
   author = {Linbo Shao and Neil Sinclair and James Leatham and Yaowen Hu and Mengjie Yu and Terry Turpin and Devon Crowe and Marko Lončar},
   doi = {10.1364/oe.397138},
   issn = {10944087},
   number = {16},
   journal = {Opt. Express},
   pages = {23728-23737},
   pmid = {32752365},
   publisher = {Optica Publishing Group},
   title = {Integrated microwave acousto-optic frequency shifter on thin-film lithium niobate},
   volume = {28},
   year = {2020},
}

@article{Zhou2024,
   author = {Yishu Zhou and Freek Ruesink and Margaret Pavlovich and Ryan Behunin and Haotian Cheng and Shai Gertler and Andrew L. Starbuck and Andrew J. Leenheer and Andrew T. Pomerene and Douglas C. Trotter and Katherine M. Musick and Michael Gehl and Ashok Kodigala and Matt Eichenfield and Anthony L. Lentine and Nils Otterstrom and Peter Rakich},
   doi = {10.1038/s41467-024-51010-8},
   journal = {Nat. Commun.},
   title = {Electrically interfaced Brillouin-active waveguide for microwave photonic measurements},
   volume = {15},
   month = {6796,},
   year = {2024},
}

@article{Bonello2001,
   author = {B. Bonello and A. Ajinou and V. Richard and Ph. Djemia and S. M. Chérif},
   doi = {10.1121/1.1399034},
   issn = {0001-4966},
   number = {4},
   journal = {J. Acoust. Soc. Am.},
   pages = {1943-1949},
   pmid = {11681374},
   publisher = {Acoustical Society of America (ASA)},
   title = {Surface acoustic waves in the GHz range generated by periodically patterned metallic stripes illuminated by an ultrashort laser pulse},
   volume = {110},
   year = {2001},
}

@article{Yu2024,
   author = {Yue Yu and Xiankai Sun},
   doi = {10.1002/lpor.202300385},
   journal = {Laser Photon. Rev. },
   month = {2300385,},
   publisher = {John Wiley and Sons Inc},
   title = {Surface Acoustic Microwave Photonic Filters on Etchless Lithium Niobate Integrated Platform},
   volume = {18},
   year = {2024},
}

@article{Snigirev2023,
   author = {Viacheslav Snigirev and Annina Riedhauser and Grigory Lihachev and Mikhail Churaev and Johann Riemensberger and Rui Ning Wang and Anat Siddharth and Guanhao Huang and Charles Möhl and Youri Popoff and Ute Drechsler and Daniele Caimi and Simon Hönl and Junqiu Liu and Paul Seidler and Tobias J. Kippenberg},
   doi = {10.1038/s41586-023-05724-2},
   issn = {14764687},
   issue = {7952},
   journal = {Nature},
   pages = {411-417},
   pmid = {36922611},
   publisher = {Nature Research},
   title = {Ultrafast tunable lasers using lithium niobate integrated photonics},
   volume = {615},
   year = {2023},
}

@article{Raphael2015,
   author = {Raphaël Van Laer and Bart Kuyken and Dries Van Thourhout and Roel Baets},
   doi = {10.1038/nphoton.2015.11},
   issn = {17494893},
   issue = {3},
   journal = {Nat. Photonics},
   pages = {199-203},
   publisher = {Nature Publishing Group},
   title = {Interaction between light and highly confined hypersound in a silicon photonic nanowire},
   volume = {9},
   year = {2015},
}

@article{Ansari2022,
   author = {Irfan Ansari and John P. George and Gilles F. Feutmba and Tessa Van De Veire and Awanish Pandey and Jeroen Beeckman and Dries Van Thourhout},
   doi = {10.1021/acsphotonics.1c01857},
   issn = {23304022},
   number = {6},
   journal = {ACS Photonics},
   month = {6},
   pages = {1944-1953},
   publisher = {American Chemical Society},
   title = {Light Modulation in Silicon Photonics by PZT Actuated Acoustic Waves},
   volume = {9},
   year = {2022},
}

@article{Balram2016,
   author = {Krishna C. Balram and Marcelo I. Davanço and Jin Dong Song and Kartik Srinivasan},
   doi = {10.1038/nphoton.2016.46},
   issn = {17494893},
   number = {5},
   journal = {Nat. Photonics},
   pages = {346-352},
   pmid = {27446234},
   publisher = {Nature Publishing Group},
   title = {Coherent coupling between radiofrequency, optical and acoustic waves in piezo-optomechanical circuits},
   volume = {10},
   year = {2016},
}

@article{Shafir2025,
   author = {Inbar Shafir and Leroy Dokhanian and Matan Slook and Shai Ben-Ami and Maayan Holsblat and Ohad Westreich and Assaf Klar and Avi Zadok},
   doi = {10.1063/5.0286374},
   journal = {APL Photonics},
   month = {096114,},
   publisher = {American Institute of Physics},
   title = {Surface acoustic wave-photonic devices in silicon nitride integrated circuits},
   volume = {10},
   year = {2025}
}

@article{Zheng2025,
   author = {Zheng Zheng and Hanke Feng and Ahmet Tarık Işık and Peter J.M. Van der Slot and Cheng Wang and David Marpaung},
   doi = {10.1515/nanoph-2025-0252},
   number = {6},
   journal = {Nanophotonics},
   volume = {14},
   pages = {4683-4690},
   title = {Gigahertz thermoelastic acousto-optic modulation in lithium niobate integrated photonic device},
   year = {2025}
}

@article{Huang2022,
   author = {Chukun Huang and Haotian Shi and Linfeng Yu and Kang Wang and Ming Cheng and Qiang Huang and Wenting Jiao and Junqiang Sun},
   doi = {10.1002/adom.202102334},
   journal = {Adv. Opt. Mater.},
   month = {2102334, },
   publisher = {John Wiley and Sons Inc},
   title = {Acousto-Optic Modulation in Silicon Waveguides Based on Piezoelectric Aluminum Scandium Nitride Film},
   volume = {10},
   year = {2022}
}

@misc{Kenning2025,
      title={Broadband acousto-optic modulators on Silicon Nitride}, 
      author={Scott E. Kenning and Tzu-Han Chang and Alaina G. Attanasio and Warren Jin and Avi Feshali and Yu Tian and Mario Paniccia and Sunil A. Bhave},
      doi = {10.1038/s41467-025-67618-3},
      year={2025},
      howpublished = {arXiv: 2505.03926},
}

@article{Freedman2025,
   author = {Jacob M. Freedman and Matthew J. Storey and Daniel Dominguez and Andrew Leenheer and Sebastian Magri and Nils T. Otterstrom and Matt Eichenfield},
   doi = {10.1038/s41467-025-65937-z},
   issn = {20411723},
   issue = {1},
   journal = {Nat. Commun. },
   month = {10959,},
   pmid = {41361157},
   publisher = {Nature Research},
   title = {Gigahertz-frequency acousto-optic phase modulation of visible light in a CMOS-fabricated photonic circuit},
   volume = {16},
   year = {2025}
}

@article{Ye2025,
   author = {Kaixuan Ye and Hanke Feng and Randy te Morsche and Chuangchuang Wei and Yvan Klaver and Akhileshwar Mishra and Zheng Zheng and Akshay Keloth and Ahmet Tarık Işık and Zhaoxi Chen and Cheng Wang and David Marpaung},
   journal = {Sci. Adv.},
   title = {Integrated Brillouin photonics in thin-film lithium niobate},
   volume = {11},
   month = {eadv4022,},
   url = {https://www.science.org},
   year = {2025},
}

@article{Marpaung2019,
   author = {David Marpaung and Jianping Yao and José Capmany},
   doi = {10.1038/s41566-018-0310-5},
   journal = {Nat. Photonics},
   pages = {80-90},
   publisher = {Nature Publishing Group},
   title = {Integrated microwave photonics},
   volume = {13},
   year = {2019}
}

@article{Wei2025,
   author = {Chuangchuang Wei and Hanke Feng and Kaixuan Ye and Maarten Eijkel and Yvan Klaver and Zhaoxi Chen and Akshay Keloth and Cheng Wang and David Marpaung},
   doi = {10.1038/s41467-025-57441-1},
   issn = {20411723},
   issue = {1},
   journal = {Nat. Commun. },
   month = {2281,},
   pmid = {40055325},
   publisher = {Nature Research},
   title = {Programmable multifunctional integrated microwave photonic circuit on thin-film lithium niobate},
   volume = {16},
   year = {2025}
}

@article{Feng2024,
   author = {Hanke Feng and Tong Ge and Xiaoqing Guo and Benshan Wang and Yiwen Zhang and Zhaoxi Chen and Sha Zhu and Ke Zhang and Wenzhao Sun and Chaoran Huang and Yixuan Yuan and Cheng Wang},
   doi = {10.1038/s41586-024-07078-9},
   issn = {14764687},
   issue = {8002},
   journal = {Nature},
   pages = {80-87},
   pmid = {38418888},
   publisher = {Nature Research},
   title = {Integrated lithium niobate microwave photonic processing engine},
   volume = {627},
   year = {2024}
}

@conference{Zheng2025-cleo,
    author = {Zheng Zheng and Ahmet Tar{i}k I\c{s}{i}k and Akshay Keloth and Peter van der Slot and David Marpaung},
    booktitle = {Conference on Lasers and Electro-Optics (CLEO)},
    title = {Thermoelastic Acoustic Emitter in Silicon Nitride Platform for Optomechanics Applications},
    year = {2025}}

@article{Bose2024,
   author = {Debapam Bose and Mark W. Harrington and Andrei Isichenko and Kaikai Liu and Jiawei Wang and Nitesh Chauhan and Zachary L. Newman and Daniel J. Blumenthal},
   doi = {10.1038/s41377-024-01503-4},
   issn = {20477538},
   issue = {1},
   journal = {Light Sci. Appl.},
   publisher = {Springer Nature},
   title = {Anneal-free ultra-low loss silicon nitride integrated photonics},
   volume = {13},
   year = {2024}
}

@article{Xiang2022,
   author = {Chao Xiang and Warren Jin and John E. Bowers},
   doi = {10.1364/prj.452936},
   issn = {23279125},
   number = {6},
   journal = {Photonics Res.},
   pages = {A82-A96},
   publisher = {Optica Publishing Group},
   title = {Silicon nitride passive and active photonic integrated circuits: trends and prospects},
   volume = {10},
   year = {2022}
}

@article{Botter2022,
   author = {Roel Botter and Kaixuan Ye and Yvan Klaver and Radius Suryadharma and Okky Daulay and Gaojian Liu and Jasper Van Den Hoogen and Lou Kanger and Peter Van Der Slot and Edwin Klein and Marcel Hoekman and Chris Roeloffzen and Yang Liu and David Marpaung},
   journal = {Sci. Adv.},
   issn = {eabq2196},
   title = {Guided-acoustic stimulated Brillouin scattering in silicon nitride photonic circuits},
   volume = {8},
   month = {eabq2196,},
   url = {https://www.science.org},
   year = {2022}
}

@article{Nejadriahi2021,
   author = {Hani Nejadriahi and Steve Pappert and Yeshaiahu Fainman and Paul Yu},
   doi = {10.1364/ol.431757},
   issn = {0146-9592},
   number = {18},
   journal = {Opt. Lett.},
   pages = {4646-4649},
   pmid = {34525072},
   publisher = {Optica Publishing Group},
   title = {Efficient and compact thermo-optic phase shifter in silicon-rich silicon nitride},
   volume = {46},
   year = {2021}
}

@article{Alemany2021,
   author = {Rubén Alemany and Pascual Muñoz and Daniel Pastor and Carlos Domínguez},
   doi = {10.3390/photonics8110496},
   journal = {Photonics},
   month = {496,},
   publisher = {MDPI},
   title = {Thermo-optic phase tuners analysis and design for process modules on a silicon nitride platform},
   volume = {8},
   year = {2021}
}

@article{Ming2010,
   author = {Ming-Chun Tien and Jared F Bauters and Martijn J R Heck and Daniel J Blumenthal and John E Bowers and J F Bauters and M J R Heck and D John and M-c Tien and A Leinse and R G Heideman and D J Blumenthal and J Bowers},
   doi = {10.3390/photonics8110496},
   journal = {Opt. Express},
   pages = {23562-23568},
   publisher = {Optica Publishing Group},
   title = {Ultra-low loss Si3N4 waveguides with low nonlinearity and high power handling capability},
   volume = {18},
   number = {23},
   year = {2010}
}

@article{Kim2017,
   author = {Sangsik Kim and Kyunghun Han and Cong Wang and Jose A. Jaramillo-Villegas and Xiaoxiao Xue and Chengying Bao and Yi Xuan and Daniel E. Leaird and Andrew M. Weiner and Minghao Qi},
   doi = {10.1038/s41467-017-00491-x},
   journal = {Nat. Commun.},
   pmid = {28851874},
   publisher = {Nature Publishing Group},
   title = {Dispersion engineering and frequency comb generation in thin silicon nitride concentric microresonators},
   volume = {8},
   month = {372},
   year = {2017}
}

@article{AghaeeRad2025,
   author = {H. Aghaee Rad and T. Ainsworth and R. N. Alexander and B. Altieri and M. F. Askarani and R. Baby and L. Banchi and B. Q. Baragiola and J. E. Bourassa and R. S. Chadwick and I. Charania and H. Chen and M. J. Collins and P. Contu and N. D’Arcy and G. Dauphinais and R. De Prins and D. Deschenes and I. Di Luch and S. Duque and P. Edke and S. E. Fayer and S. Ferracin and H. Ferretti and J. Gefaell and S. Glancy and C. González-Arciniegas and T. Grainge and Z. Han and J. Hastrup and L. G. Helt and T. Hillmann and J. Hundal and S. Izumi and T. Jaeken and M. Jonas and S. Kocsis and I. Krasnokutska and M. V. Larsen and P. Laskowski and F. Laudenbach and J. Lavoie and M. Li and E. Lomonte and C. E. Lopetegui and B. Luey and A. P. Lund and C. Ma and L. S. Madsen and D. H. Mahler and L. Mantilla Calderón and M. Menotti and F. M. Miatto and B. Morrison and P. J. Nadkarni and T. Nakamura and L. Neuhaus and Z. Niu and R. Noro and K. Papirov and A. Pesah and D. S. Phillips and W. N. Plick and T. Rogalsky and F. Rortais and J. Sabines-Chesterking and S. Safavi-Bayat and E. Sazhaev and M. Seymour and K. Rezaei Shad and M. Silverman and S. A. Srinivasan and M. Stephan and Q. Y. Tang and J. F. Tasker and Y. S. Teo and R. B. Then and J. E. Tremblay and I. Tzitrin and V. D. Vaidya and M. Vasmer and Z. Vernon and L. F.S.S.M. Villalobos and B. W. Walshe and R. Weil and X. Xin and X. Yan and Y. Yao and M. Zamani Abnili and Y. Zhang},
   doi = {10.1038/s41586-024-08406-9},
   journal = {Nature},
   pages = {912-919},
   pmid = {39843755},
   publisher = {Nature Research},
   title = {Scaling and networking a modular photonic quantum computer},
   volume = {638},
   year = {2025}
}

@article{Zhou2022sci,
   author = {Junchao Zhou and Diana Al Husseini and Junyan Li and Zhihai Lin and Svetlana Sukhishvili and Gerard L. Coté and Ricardo Gutierrez-Osuna and Pao Tai Lin},
   doi = {10.1038/s41598-022-09597-9},
   issn = {20452322},
   issue = {5572},
   journal = {Scientific Reports},
   pmid = {35368033},
   publisher = {Nature Research},
   title = {Detection of volatile organic compounds using mid-infrared silicon nitride waveguide sensors},
   volume = {12},
   month = {5572,},
   year = {2022}
}

\newpage
\onecolumngrid
\beginsupplement
\newpage




\end{document}